\documentclass{article}
\usepackage{spconfa4}
\usepackage{csquotes}
\usepackage{amsmath,amssymb,mathtools,bm}
\usepackage{siunitx}
\usepackage{booktabs}
\usepackage{multirow}
\usepackage{graphicx}
\usepackage{enumitem}
\usepackage{url}
\usepackage{balance}

\title{Masked Autoregressive Speech Enhancement \\ with Continuous Neural Audio Codec Representations}

\name{Yoto Fujita$^{1,2}$ $\quad$ Simon Leglaive$^{1}$ $\quad$  Laurent Girin$^{2}$}
\address{$^{1}$ CentraleSup\'elec, IETR (UMR CNRS 6164), France\\
$^{2}$ Univ. Grenoble Alpes, CNRS, Grenoble-INP, GIPSA-lab, France}

\begin{document}
\ninept
\maketitle

\begin{abstract}
Most previous work on speech enhancement (SE) based on masked generative modeling relied on discrete token representations of audio signals, obtained using neural audio codecs (NACs). However, a recent study has shown that continuous latent representations of NACs can be advantageous for SE in terms of speech quality and intelligibility. In this work, we propose masked autoregressive SE (MARSE), a method for SE based on iterative decoding of masked clean speech frames using continuous NAC representations of speech. In particular, we investigate a set of different decoding policies, \textit{ceteris paribus}, that is, using the same DNN (a Conformer model), the same NAC (the DAC codec) and the same training setup. The results show that MARSE enables a flexible trade-off between SE performance and computational cost. Audio examples and code are available online.\footnote{\url{https://yotofujita.github.io/marse}. This research was supported by ANR projects ANR-24-CE23-2229 and ANR-23-CE23-0009.}
\end{abstract}


\section{Introduction}
\label{sec:intro}

Speech enhancement (SE) in noise has a long history of conventional signal processing techniques \cite{benesty2006speech, loizou2013speech} and has been largely revisited in the last decade with deep neural networks (DNNs) \cite{wang2018supervised}. DNN-based SE has been increasingly performed in learned latent spaces instead of using conventional representations such as the short-time Fourier transform. In particular, the compact yet expressive discrete latent representations provided by neural audio codecs (NAC), which consist of sequences of vector quantizer codes called tokens, have recently facilitated SE based on token sequence modeling with Transformers \cite{vaswani2017attention}, in the line of speech language models~\cite{mousaviDiscreteAudioTokens2025b}. The tokens corresponding to a clean speech signal frame at time $t$ are decoded (i.e.,~predicted) using the noisy signal and possibly previously decoded speech tokens. 
Token-based SE has been explored with different decoding strategies: non-autoregressive decoding~\cite{kammounModelingStrategiesSpeech2026,wangSELMSpeechEnhancement2024a}, causal autoregressive decoding~\cite{kammounModelingStrategiesSpeech2026,yaoGenSEGenerativeSpeech2024a,xueLowLatencySpeechEnhancement2024a}, or non-causal masked generative decoding~\cite{yangGenhancerHighFidelitySpeech2024b,liMaskSRMaskedLanguage2024b,zhangAnyEnhanceUnifiedGenerative2025,phamMAGECoarsetoFineSpeech2026}. 

This paper aims at both investigating SE models that operate on \textit{continuous} NAC representations and extending the study of decoding policies for SE. This is motivated by the following three points. First, although token-based SE methods are reported to exhibit good speech quality, they can suffer from limited preservation of acoustic details that matter for speech quality and intelligibility, leading to reduced performance on some downstream tasks such as automatic speech recognition (ASR)~\cite{yangGenhancerHighFidelitySpeech2024b}. Second, it has been recently shown in \cite{kammounModelingStrategiesSpeech2026} that using a continuous NAC representation---that is, embedding vectors at the output of the NAC encoder and before the quantizer---can substantially improve speech intelligibility and quality. Third, most of the token-based methods mentioned above focus on a specific decoding policy with a specific experimental setup, and the trade-off between performance and computational cost regarding decoding policy has not been explored. 

\begin{figure}[tbp]
\centering
\includegraphics[width=0.49\textwidth]{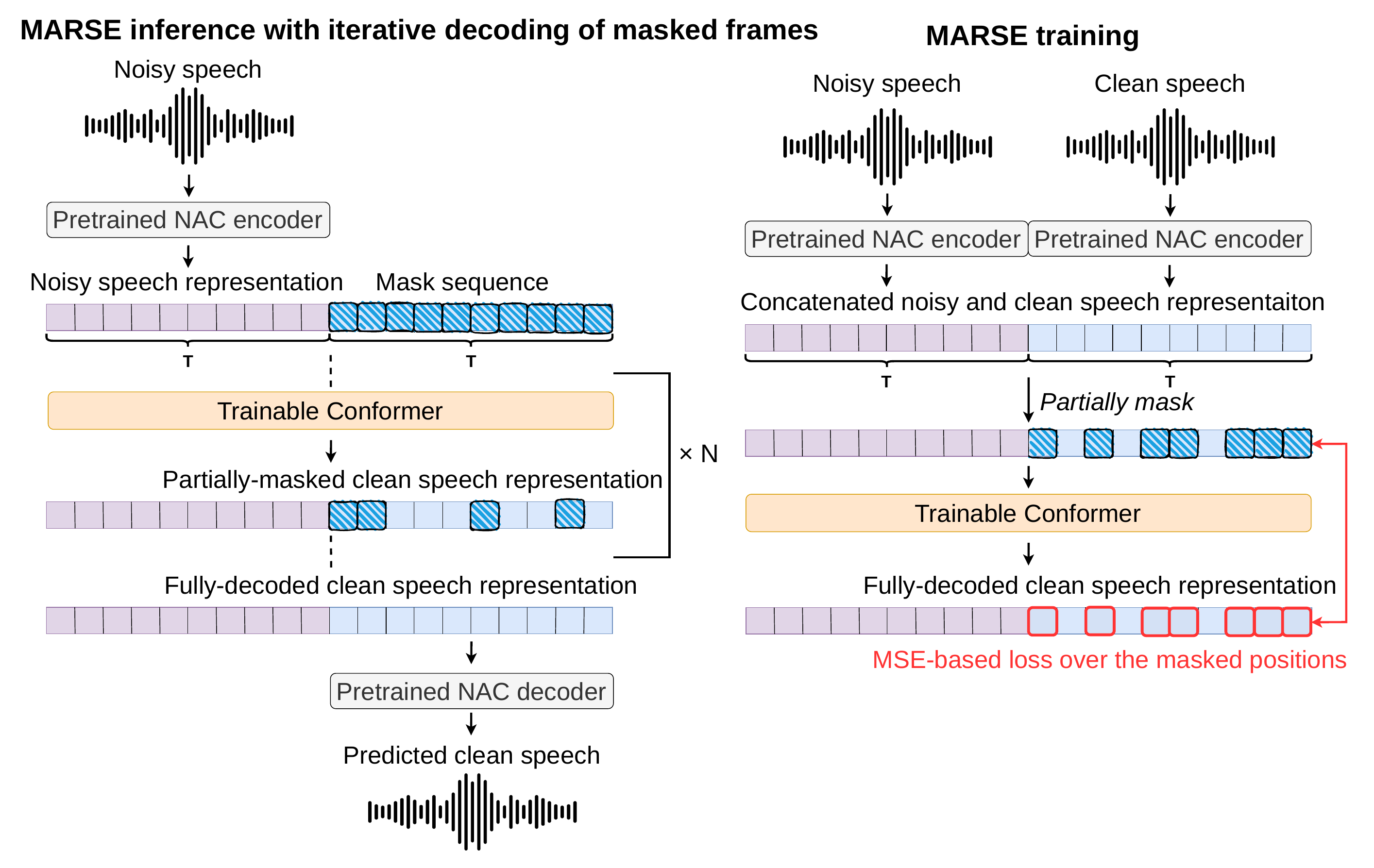}\caption{Overview of the proposed MARSE method applied on a continuous NAC representation (left: inference; right: training). }
\label{fig:mgse_overview}
\end{figure}

Independently of the SE problem, a unified framework called masked autoregressive (MAR) modeling was recently proposed for image generation in \cite{liAutoregressiveImageGeneration2024a}. There, decoding is formulated as an iterative unmasking process, providing a unified view of autoregressive (AR) and masked generation. This framework was then adapted for SE with continuous representations in \cite{yang2025investigating}. In the present paper, we expand the work of  \cite{kammounModelingStrategiesSpeech2026}, \cite{liAutoregressiveImageGeneration2024a}, and \cite{yang2025investigating} and propose masked autoregressive speech enhancement (MARSE), in which iterative decoding is formulated as an AR probabilistic process over \textit{blocks} of continuous NAC representations. Our objective is not to introduce a new decoding approach, but rather to provide a unified framework encompassing AR, non-AR and masked AR decoding policies. With a particular focus on the role and impact of decoding policies, our work offers complementary insights to previous work. 
Concretely, we compare several decoding policies as variants of the number of iterations and/or the definition of the blocks. We study their effect on speech quality, intelligibility, and computational cost, by a fair comparison using the same network architecture (a Conformer-based model \cite{gulatiConformerConvolutionaugmentedTransformer2020a}), the same NAC (Descript Audio Codec \cite{kumarHighFidelityAudioCompression2023d}), and the same training setup. 
Experiments show that the proposed MARSE method allows for a flexible trade-off between SE performance and computational cost.

\section{Method}
\label{sec:method}

In a probabilistic framework, SE can be defined as the problem of estimating a conditional distribution $p_\theta(\mathbf{x}\!\mid\!\mathbf{y})$ of parameters $\theta$, where $\mathbf{x}, \mathbf{y} \in \mathcal{A}$ denote the clean and noisy speech signals respectively, defined in some representation domain $\mathcal{A}$. Inspired by \cite{kammounModelingStrategiesSpeech2026}, we consider a model that operates on the continuous latent representation of a pretrained NAC, before quantization. We thus have $\mathbf{x} = \{ \mathbf{x}_t \in \mathbb{R}^D \}_{t=1}^T$ and $\mathbf{y} = \{ \mathbf{y}_t \in \mathbb{R}^D \}_{t=1}^T$, where $T$ denotes the number of time frames and $D$ the NAC's latent space dimension. Deep learning-based SE methods rely on DNNs to parametrize the distribution $p_\theta(\mathbf{x}\!\mid\!\mathbf{y})$. In a supervised setting, the parameters $\theta$ are learned using a labeled dataset of paired clean and noisy speech signals, typically by minimizing the negative log-likelihood $- \ln p_\theta(\mathbf{x}~\mid~\mathbf{y})$ averaged over the training examples.

\subsection{Masked autoregressive speech enhancement}

Within this probabilistic framework, defining $p_\theta(\mathbf{x}\!\!\mid\!\!\mathbf{y})$ involves (i) defining the distribution of the clean speech variables $\mathbf{x}_t$, $t \in \{1,...,T\}$, and their temporal dependencies; and (ii) defining the DNN architecture that implements this probabilistic model. In this work, we apply the MAR framework of~\cite{liAutoregressiveImageGeneration2024a} to the SE problem, resulting in the class of MARSE models, which allows for arbitrary assumptions about these dependencies and tackles SE as an iterative decoding process, as illustrated in the left part of Fig.~\ref{fig:mgse_overview}. In MARSE, the conditional distribution of $\mathbf{x}$ given $\mathbf{y}$ is defined by:\footnote{The diffusion model used in the original MAR framework \cite{liAutoregressiveImageGeneration2024a} is replaced here by a simpler, and likely less effective, Gaussian model. Although this simplification may affect the overall performance, it does not impact the focus of this study, which is to formalize and compare different decoding policies.}
\begin{align}
p_\theta (\mathbf{x} \mid \mathbf{y}) &= \prod\nolimits_{i=1}^N p_\theta \bigl( \mathbf{x}_{M(i)} \mid \mathbf{y}, \mathbf{x}_{V(i)} \bigr) \label{eq:cond_dist_factorization} \\
&= \prod\nolimits_{i=1}^N \prod\nolimits_{t \in M(i)} \mathcal{N}(\mathbf{x}_t; f_{\theta, t}(\mathbf{y}, \mathbf{x}_{V(i)}), \mathbf{I}), \label{eq:cond_dist_factorization_gaussian}
\end{align}
where $N \in \{1,...,T\}$ is the number of iterations in the decoding process and
\begin{itemize}[leftmargin=*]
    \item $\mathbf{x}_{M(i)}$ denotes the set of predicted frames at iteration $i$, with $M(i) \subseteq \{1,...,T\}$ and such that $\bigcup_{i=1}^{N} M(i) = \{1,...,T\}$;
    \item $\mathbf{x}_{V(i)}$ denotes the set of visible (i.e.~previously decoded) frames at the beginning of iteration $i$, with $V(1) = \emptyset$, $V(i+1) = V(i) \, \cup \, M(i)$ and $M(i) \cap V(i) = \emptyset$ for $i \ge 1$.
    \item and $f_{\theta} : \mathbb{R}^{D \times T} \times \mathbb{R}^{D \times T} \mapsto \mathbb{R}^{D \times T}$ denotes a neural network that processes the noisy speech $\mathbf{y}$ and visible clean speech frames $\mathbf{x}_{V(i)}$ to predict $\mathbf{x}_{M(i)}$. Inspired by \cite{kammounModelingStrategiesSpeech2026}, we use a Conformer-based network taking as input the temporal concatenation of noisy and partially-masked clean speech representations. The masked clean speech frames at positions $t \in \{1,...,T\} \setminus V(i)$ are here replaced by a learnable mask vector.
\end{itemize}
By construction, the collection of index sets $\{M(i)\}_{i=1}^N$ is a disjoint partition of $\{1,...,T\}$. The factorization in \eqref{eq:cond_dist_factorization} is therefore an application of the chain rule of probability that constitutes an \emph{autoregressive model over blocks of frames}, which forms a proper probability density function (PDF) over $\mathbf{x}$ as long as each conditional factor in \eqref{eq:cond_dist_factorization} is also a valid PDF. Eq.~\eqref{eq:cond_dist_factorization_gaussian} shows that within a block, the vectors $\mathbf{x}_t$ are assumed to be mutually conditionally independent.

\subsection{Decoding policies for inference}
\label{subsec:decoding-policies}

In MARSE, a fundamental aspect of the model is the definition of the collection of index sets $\{M(i)\}_{i=1}^N$, which will be referred to as the \emph{decoding policy} in the following sections. This policy directly relates to the assumed temporal dependencies between the clean speech frames. The flexibility of MARSE comes from the possibility of designing various decoding policies and, in particular, choosing an arbitrary number of iterations $N \in \{1,...,T\}$. Below, we present causal and non-causal decoding policies, which can all be characterized in two steps: (i) defining the number of frames decoded in parallel at iteration $i$, which is equal to the cardinality of $M(i)$ in \eqref{eq:cond_dist_factorization}, denoted by $\mathrm{card}(M(i))$; (ii) defining the indices of the frames to predict, that is, specifying the set $M(i)$ given its cardinality. In this work, following \cite{changMaskGITMaskedGenerative2022a}, the cardinality of $M(i)$ is fixed according to a predefined cosine schedule. More precisely, $\mathrm{card}(M(i)) = n_{i+1} - n_{i}$ frames, where ${n_i = \mathrm{card}(V(i))} = \left\lfloor T \left( 1 - \gamma(\frac{i-1}{N}) \right) \right\rfloor$, with $\gamma : r \in [0, 1] \mapsto \cos\left( \frac{\pi}{2} r \right) \in [0,1]$. Here, $n_i$ denotes the number of predicted frames up to, and not including, iteration $i $, such that $n_1 = \mathrm{card}(V(1)) = 0$. Then, variants of the decoding policy are obtained by specifying different strategies for defining $M(i)$ at each iteration and/or by choosing a different number of iterations.

\vspace{0.15cm}

\noindent \textbf{Causal decoding.} In the block-wise causal AR decoding policy, the set of indices to predict is defined by $M(i) = \{n_i + 1, n_i + 2, ..., n_{i+1}\}$. 
Thus, each block of frames consists of a subset of consecutive frames, and the blocks follow a causal ordering along the iterations. 
In the limit case where $N=1$, we have $M(1) = \{1,...,T\}$ and $V(1) = \emptyset$, i.e.~all frames are decoded all at once and the decoding is not really AR anymore (we thus can refer to it as non-AR). As can be seen from \eqref{eq:cond_dist_factorization_gaussian}, this non-AR decoding assumes that all variables $\mathbf{x}_t$, $t \in \{1,...,T\}$, are mutually independent. It constitutes the most naive temporal model.
In the limit case where $N = T$, we recover the conventional frame-wise causal AR decoding with $M(i) = \{i\}$ and $V(i) = \{1,..., i-1\}$ for all $i \in \{1,...,T\}$. This policy is the most expressive in terms of temporal modeling, as no conditional independence assumptions are made. However, it requires $T$ forward passes in the neural network $f_{\theta}$, which can be computationally expensive.

\vspace{0.15cm}

\noindent \textbf{Non-causal decoding with oracle index selection.} Contrary to the previous policy, in the non-causal decoding policy a masked frame at a given time step can be decoded conditioned on frames in the future. The intuition behind a non-causal policy for SE is that frames associated with high signal-to-noise ratio can be decoded first, and then used to decode noisier frames from the past. In the non-causal decoding policy with oracle index selection, at iteration $i$, we first decode all the frames that remain to be decoded. Then, we use the ground-truth clean speech to select, among those frames, the $\mathrm{card}(M(i))$ ones that have the lowest prediction error (in terms of squared error). This decoding policy is not realistic, as it requires the availability of the ground-truth clean speech, but it serves as an oracle baseline for evaluating the performance of MARSE with non-causal decoding.

\vspace{0.15cm}

\noindent \textbf{Non-causal decoding with random index selection.} We also consider the most naive non-causal decoding policy based on random index selection, as used in MAR models \cite{liAutoregressiveImageGeneration2024a}. In this policy, given its cardinality, at each iteration $i \in \{1,...,N\}$ the set $M(i)$ is filled with indices chosen randomly within the set of remaining masked frames $\{1,...,T\}\setminus V(i)$, following a uniform distribution. 
This policy can be interpreted as randomly partitioning $\{1,...,T\}$ into $N$ blocks of size defined by the cosine schedule function $\gamma$, and modeling the probabilistic dependencies between these blocks with \eqref{eq:cond_dist_factorization}.

\subsection{Training}
\label{subsec:training}

The training of the MARSE model is illustrated in the right part of Fig.~\ref{fig:mgse_overview}. Given a labeled dataset of noisy-clean speech pairs $(\mathbf{y}, \mathbf{x})$, it is done by optimizing the negative log-likelihood defined from \eqref{eq:cond_dist_factorization_gaussian} (averaged over the training set). This is equivalent to minimizing the following MSE-based loss function:
\begin{equation}
\mathcal{L}(\theta; \mathbf{x}) = \mathbb{E}_{\mathcal{U}(i; \{1, ..., N\})} \bigg[ \sum\limits_{t \in M(i)} \left\| \mathbf{x}_t - f_{\theta, t}\left(\mathbf{y}, \mathbf{x}_{V(i)}\right) \right\|_2^2 \bigg],
\label{eq:loss}
\end{equation}
where summation over $i \in \{1,...,N\}$ has been equivalently replaced by an expectation. In masked generative modeling, it is common to approximate the expectation using one single sample $i$ \cite{changMaskGITMaskedGenerative2022a}. We follow this approach, where the number of masked frames is computed using the same cosine schedule as defined before for the decoding policies, i.e.~$\mathrm{card}(M(i)) = \left\lfloor \gamma(r) \cdot T \right\rfloor$ where $r = (i-1)/N$. In practice, $r$ is sampled from $\mathcal{U}([0, 1[)$, which is consistent with $r = (i-1)/N$ and $i \sim \mathcal{U}(\{1, ..., N\})$ as $N \rightarrow + \infty$.
Given $\mathrm{card}(M(i))$, the set $M(i)$ is then defined randomly within $\{1,...,T\}$, following a uniform distribution. Finally, $V(i)$ is set to $\{1,...,T\} \setminus M(i)$.

\subsection{Mitigating exposure bias with quantization}
\label{sec:exposure_bias}

At inference, for each iteration $i \in \{1,...,N\}$ of the decoding process, we compute the prediction of the clean speech frames $\mathbf{x}_{M(i)} = \{ f_{\theta, t}(\mathbf{y}, \mathbf{x}_{V(i)}) \}_{t \in M(i)}$, which corresponds to the mean of the Gaussian distributions in \eqref{eq:cond_dist_factorization_gaussian}. In the next iteration, these predicted frames are reinjected into the model as visible frames: $\mathbf{x}_{V(i+1)} = \mathbf{x}_{V(i)} \cup \mathbf{x}_{M(i)}$  with $\mathbf{x}_{V(1)} = \emptyset$. This is different from the training process, where the frames $\mathbf{x}_{V(i)}$ in the loss function \eqref{eq:loss} are obtained from ground-truth clean speech. This mismatch between training and testing conditions, commonly referred to as exposure bias \cite{ranzato2016sequence}, can result in error accumulation during decoding iterations. To mitigate this problem, as proposed in \cite{kammounModelingStrategiesSpeech2026}, we leverage the fact that we work in the latent space of a NAC to quantize the visible frames $\mathbf{x}_{V(i)}$ using the pretrained NAC quantizer, before feeding them to the model $f_{\theta}$. This is done both at training and inference.

\begin{table*}[t]
\centering
\small
\resizebox{.8\linewidth}{!}{ 
\begin{tabular}{lccccc|ccccc} 
\toprule
& \multicolumn{5}{c}{\textbf{Libri1Mix}} 
& \multicolumn{5}{c}{\textbf{LibriDEMAND}} \\
\toprule
\textbf{Model} 
& \textbf{SIG} $\uparrow$ 
& \textbf{BAK} $\uparrow$ 
& \textbf{OVRL} $\uparrow$ 
& \textbf{dWER} $\downarrow$ 
& \textbf{GFLOPs} $\downarrow$ 
& \textbf{SIG} $\uparrow$ 
& \textbf{BAK} $\uparrow$ 
& \textbf{OVRL} $\uparrow$ 
& \textbf{dWER} $\downarrow$ 
& \textbf{GFLOPs} $\downarrow$ \\
\midrule
Noisy & 2.46 & 1.81 & 3.08 & 30.19 & - & 2.55 & 2.10 & 1.92 & 21.77 & - \\
\midrule
ConvTasNet~\cite{luoConvTasNetSurpassingIdeal2019a} & 3.47 & \underline{4.10} & 3.22 & \textbf{9.79} & 49 & 3.41 & \bf 3.85 & 3.02 & 10.49 & 51 \\
DPTNet~\cite{chenDualPathTransformerNetwork2020} & 3.44 & 4.08 & 3.16 & \underline{10.05} & 12 & 3.38 & \underline{3.84} & 2.98 & 9.57 & 12 \\
\midrule
C-NAR \cite{kammounModelingStrategiesSpeech2026} & 3.60 & 4.08 & 3.32 & 12.84 & 1235 & \underline{3.55} & 3.76 & 3.08 & 9.61 & 1334 \\
C-AR \cite{kammounModelingStrategiesSpeech2026} & \bf 3.64 & \bf 4.11 & \bf 3.37 & 20.89 & 3856 & \underline{3.55} & 3.76 & 3.09 & 15.91 & 4166 \\
\midrule
MARSE-causal & \underline{3.62} & \underline{4.10} & \underline{3.34} & 12.68 & 1912 & 3.54 & 3.80 & 3.11 & \underline{9.35} & 2065 \\
MARSE-NC-random & \underline{3.62} & 4.09 & \underline{3.34} & 13.39 & 1912 & 3.54 & 3.83 & \underline{3.12} & 10.13 & 2065 \\
MARSE-NC-oracle & 3.61 & 4.08 & 3.34 & 12.87 & 1912 & \bf 3.57 & 3.83 & \bf 3.14 & \bf 8.86 & 2065 \\
\bottomrule
\end{tabular}
}
\caption{Performance obtained by the proposed MARSE model for $N=10$ iterations and the baselines on the Libri1Mix in-domain dataset and LibriDEMAND out-of-domain dataset. 
The Metrics are described in the text. The scores for input noisy speech (`Noisy' item) are provided as a lower reference. Best and second-best scores in each column are bold and underlined.}
\label{tab:librimix}
\vspace{-0.3cm}
\end{table*}

\section{Experiments}
\label{sec:exp}

This section describes the experiments we conducted to assess MARSE on continuous NAC representations with the different decoding policies, under the same general configuration (in terms of NAC representation, dataset, DNN architecture, training setup, and evaluation metrics). 

\subsection{Experimental setup}
\label{subsec:experimental_setup}

\noindent \textbf{NAC representation.} 
Regarding the NAC representation, we used DAC~\cite{kumarHighFidelityAudioCompression2023d}, with continuous latent representations of dimension $D = 1024$. 
DAC is one of the most widely used NACs, based on a convolutional encoder-decoder architecture and a 12-stage residual vector quantizer (RVQ), whose codebooks each contain $1024$ codewords.
\vspace{0.1cm}

\noindent \textbf{Data.} 
For MARSE training and validation, we used the 212-hour $\texttt{train-360}$ and 11-hour $\texttt{dev}$ subsets of Libri1Mix, respectively. This dataset is obtained by discarding one of the two speakers in Libri2Mix mixtures~\cite{cosentinoLibriMixOpenSourceDataset2020a}. 
Each subset of Libri1Mix contains paired noisy and clean utterances at 16~kHz
generated by mixing clean speech from LibriSpeech~\cite{panayotovLibrispeechASRCorpus2015} with noise from WHAM!~\cite{wichernWHAMExtendingSpeech2019a}. 
We used the 11-hour $\texttt{test}$ subset of Libri1Mix to evaluate in-domain performance. 
To further evaluate out-of-domain performance, we created a 4-hour synthetic dataset, here referred to as LibriDEMAND, by mixing clean speech from the $\texttt{train-clean-100}$ subset of LibriSpeech with noise from DEMAND~\cite{thiemannDiverseEnvironmentsMultichannel2013a} with a very similar procedure to Libri1Mix.
For training, a 1-second clip is randomly cropped from each utterance to form a sequence of embedding vectors with length $T=50$.
At inference, longer utterances are segmented into 1~s-segments that are processed independently by the model.

\vspace{0.1cm}

\noindent \textbf{DNN architecture.} We used a Conformer architecture~\cite{gulatiConformerConvolutionaugmentedTransformer2020a} with 16 Conformer blocks, hidden dimension set to 384, 12 attention heads, each of dimension 32, convolution kernel of size 10, expansion factor for convolution module set to 2, and expansion factor for feed-forward module set to 4. To align the dimension of DAC embedding vectors with the hidden dimension of the Conformer, two learnable linear layers are applied before and after the Conformer.

\vspace{0.1cm}

\noindent \textbf{Training setup.} The model was trained by multi-GPU batch gradient descent, using the AdamW optimizer~\cite{loshchilovDecoupledWeightDecay2018} with a learning rate of $10^{-3}$, beta coefficients of $(0.9, 0.95)$, and weight decay of $0.05$.
The batch size was set to $128$ for each GPU and the number of epochs was fixed to $300$.
The learning rate was linearly increased from 0 to the target value over 10 epochs and followed a cosine decay schedule over the remaining $290$ epochs.
The multi-GPU training was implemented with the distributed data parallel module of PyTorch across four NVIDIA RTX A100 GPUs with 40GB VRAM.

\vspace{0.1cm}

\noindent \textbf{Evaluation metrics.} Since NAC outputs are not aligned sample-wise with the clean speech reference due to adversarial training, following prior works~\cite{kammounModelingStrategiesSpeech2026,wangSELMSpeechEnhancement2024a}, the enhanced speech quality was measured using the non-intrusive DNSMOS P.835 SIG, BAK and OVRL scores~\cite{reddyDnsmosP835NonIntrusive2022a} (between $1.0$ and $5.0$, the higher the better), which measure respectively the quality of the speech signal, the intrusiveness of background noise, and the overall quality.
Speech intelligibility was measured by an ASR-based proxy for phonetic preservation, which is the differential word error rate (dWER in $\%$, the lower the better) between the transcriptions obtained from the enhanced and ground-truth speech. We used a pretrained ASR model based on wav2vec 2.0~\cite{baevskiWav2vec20Framework2020a}.\footnote{https://huggingface.co/facebook/wav2vec2-base-960h}
As for computational cost analysis, we measured the amount of giga floating-point operations (GFLOPs), including the encoding and decoding by DAC.

\vspace{0.1cm}

\noindent \textbf{Baselines.} As conventional non-AR SE baselines, we used ConvTasNet~\cite{luoConvTasNetSurpassingIdeal2019a} and DPTNet~\cite{chenDualPathTransformerNetwork2020}, using the publicly-available models pretrained on Libri1Mix.\footnote{https://huggingface.co/JorisCos} We also compare MARSE against two NAC-based SE models from \cite{kammounModelingStrategiesSpeech2026} that we retrained. Those do not use masking at training or inference, although they have a specific decoding policy: (i) C-NAR is a non-AR model using the `all at once' decoding corresponding to $N=1$, and (ii) C-AR is a frame-wise causal AR model, corresponding to $N=T$. 
Concretely, C-NAR was trained and inferred using a single-step noisy-to-clean mapping, whereas C-AR was trained with a next-frame prediction objective and inferred via frame-wise causal AR decoding of clean representations. Note that we do not use token-based baselines, as it has been shown in \cite{kammounModelingStrategiesSpeech2026} that C-NAR and C-AR significantly outperform their counterparts using token representations.

\begin{figure}[tbp]
\centering
\includegraphics[width=0.4\textwidth]{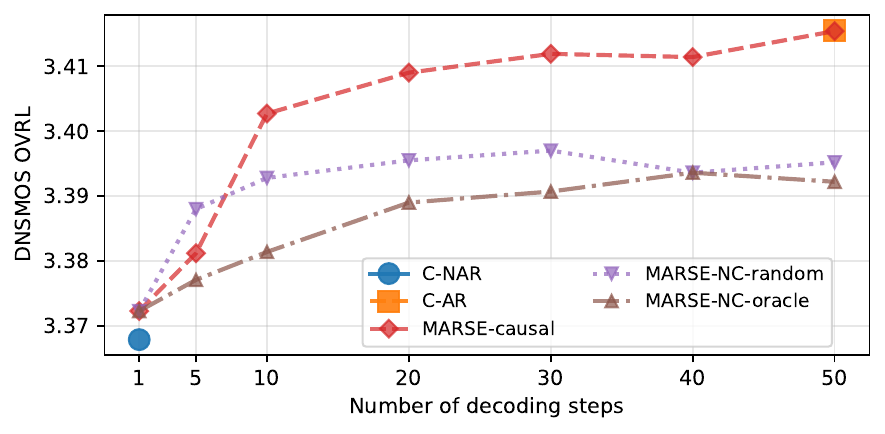}
\caption{DNSMOS OVRL score obtained by the proposed MARSE model (for the three decoding policies) and the C-AR and C-NAR baselines on a subset of 300 samples from Libri1Mix $\texttt{test}$, as a function of the number of decoding iterations.}
\label{fig:mgse_results_vs_steps_librimix}
\vspace{-0.5cm}
\end{figure}

\vspace{-0.25cm}
\subsection{Results}

Table~\ref{tab:librimix} provides the results obtained by the proposed MARSE model with the three decoding policies presented in Section~\ref{subsec:decoding-policies}, using $N=10$ decoding steps, as well as those obtained by the baselines. Causal decoding is denoted `MARSE-causal'. Non-causal decoding with random index selection is denoted `MARSE-NC-random' and constitutes the practical non-causal baseline. Non-causal decoding with oracle index selection is denoted `MARSE-NC-oracle' and is reported as an oracle baseline.  
Regarding the non-AR baselines, ConvTasNet and DPTNet, the SE quality measured by SIG and OVRL was globally lower than for the other models, but their computational cost was very reasonable. The C-AR and C-NAR baselines showed contrasting performance on the in-domain Libri1Mix. 
C-AR exhibited higher quality (SIG = $3.64$, BAK = $4.11$, OVRL = $3.37$) but slower inference (GFLOPs = $3856$), while C-NAR exhibited lower quality (SIG = $3.60$, BAK = $4.08$, OVRL = $3.32$) but faster inference (GFLOPs =  $1235$). 
This observation is consistent with the assumption behind these models: C-AR fully models the causal dependency of speech frames at the expense of computational cost, as it requires $T$ forward passes in the model for inference, while C-NAR independently models each frame for the benefit of computational cost, as it requires a single forward pass in the model for inference.
In general, the MARSE model, which assumes inter-block dependency but inner-block independence, demonstrated intermediate SE performance between C-NAR and C-AR, and its computational cost was also in between (e.g., SIG = $3.62$, BAK = $4.10$, OVRL = $3.34$,  GFLOPs = $1912$ for MARSE-causal). This provides an illustration of the flexibility of the proposed MARSE method in terms of trade-off between SE performance and computational cost.

Another noticeable result shown in Table~\ref{tab:librimix} is that the MARSE models exhibited intelligibility clearly better than C-AR and even competitive with C-NAR on both the in-domain Libri1Mix (e.g., dWER is $12.68$\% for MARSE-causal, $20.89$\% for C-AR, and $12.84$\% for C-NAR) and out-of-domain LibriDEMAND (e.g., dWER is $9.35$\% for MARSE-causal, $15.91$\% for C-AR, and $9.61$\% for C-NAR).
Among the three MARSE models, the practical MARSE-causal performed the best on in-domain data, whereas the non-practical MARSE-NC-oracle exhibited the best scores on out-of-domain data. This is likely because MARSE-causal better models the temporal dependencies in clean speech on in-domain data, while accumulated errors across decoding iterations are more pronounced on out-of-domain data, making a more reliable decoding policy necessary. 
These results suggest that better non-causal frame-selection strategies could be worth exploring in future work.

Fig.~\ref{fig:mgse_results_vs_steps_librimix} displays the DNSMOS OVRL score obtained by the proposed MARSE model (for the three decoding policies) as a function of the number of decoding iterations $N \in \{1, 5, 10, 20, 30, 40,50\}$. 
To limit computational cost and duration of the experiments, these scores were evaluated on a subset of 300 samples randomly selected from Libri1Mix $\texttt{test}$, therefore they can be different from those of Table~\ref{tab:librimix}. 
Even though the absolute score difference among decoding policies remains moderate, we can observe in Fig.~\ref{fig:mgse_results_vs_steps_librimix} an improvement in SE quality as the number of decoding steps increases.
In particular, it is noticeable that the OVRL score of MARSE-causal closely approaches that of C-NAR for $N=1$ and that of C-AR for $N=50$.
This result is consistent with the fact that for $N=1$, both MARSE and C-NAR employ `all-at-once' frame decoding, and for $N=50$, both MARSE-causal and C-AR employ frame-wise causal AR decoding. 
Between these two `extreme' decoding policies, the MARSE-causal performance improves between C-NAR and C-AR as the number of iterations increases. 
This confirms the flexibility of the MARSE framework for exploring variants of decoding policies and enabling a trade-off between SE performance and computational cost. 
In particular, we see in Fig.~\ref{fig:mgse_results_vs_steps_librimix} that the performance of MARSE-causal starts to stagnate in the range $20$--$40$ iterations. Setting $N$ in this range would thus provide a good trade-off between performance and computational cost. 
Note that all MARSE models were trained with the exact same strategy presented in Section~\ref{subsec:training}.

\vspace{-0.3cm}
\section{Conclusion}
\label{sec:conclusion}
\vspace{-0.2cm}

In this work, we proposed the MARSE method and explored different iterative decoding policies for SE using continuous NAC representations. 
The experimental results showed the ability of this method to provide a flexible trade-off between SE performance and computational cost. 
Future work may focus on developing a non-oracle confidence measure to determine a more effective non-causal decoding order than random index selection, which may be beneficial in terms of generalization, as suggested by the results obtained with MARSE-NC-oracle on the out-of-domain data.

\balance
\bibliographystyle{IEEEtran}
\bibliography{refs}

\end{document}